\documentclass[superscriptaddress,secnumarabic,
amssymb,amsmath,nobibnotes,aps,prd,showkeys,showpacs,nofootinbib]{revtex4}%
\usepackage{graphicx}
\usepackage{epsf}
\usepackage{bm}
\usepackage{amsmath}
\usepackage{amsfonts}
\usepackage{amssymb}
\usepackage{epstopdf}
\usepackage{natbib}
\usepackage{color}
\usepackage{float}
\providecommand{\U}[1]{\protect\rule{.1in}{.1in}}

\newcommand{\be}{\begin{equation}}
\newcommand{\ee}{\end{equation}}

\newcommand{\mincir}{\raise
-3.truept\hbox{\rlap{\hbox{$\sim$}}\raise4.truept\hbox{$<$}\ }}
\newcommand{\magcir}{\raise
-3.truept\hbox{\rlap{\hbox{$\sim$}}\raise4.truept\hbox{$>$}\ }}

\newtheorem{remark}{Remark}[section]

\begin{document}

\title{Gravitational particle production of superheavy massive particles in Quintessential Inflation: A numerical analysis}

%\author{Supriya Pan}
%\email{supriya.maths@presiuniv.ac.in}
%\affiliation{Department of Mathematics, Presidency University, 86/1 College Street, 
%Kolkata 700073, 
%India}

\author{Llibert Arest\'e Sal\'o}
\email{l.arestesalo@qmul.ac.uk} 
%\affiliation{Departament de Matem\`atiques, Universitat Polit\`ecnica de Catalunya, Diagonal 647, 08028 Barcelona, Spain}
\affiliation{School of Mathematical Sciences, Queen Mary University of London, Mile End Road, London, E1 4NS, United Kingdom}

\author{Jaume  de Haro}
\email{jaime.haro@upc.edu}
\affiliation{Departament de Matem\`atiques, Universitat Polit\`ecnica de Catalunya, Diagonal 647, 08028 Barcelona, Spain}

\thispagestyle{empty}

\begin{abstract}
We compute numerically the reheating temperature due to the gravitational production of conformally coupled superheavy  particles during the  phase transition from the end of inflation to the beginning of  kination in two different  Quintessential Inflation (QI) scenarios, namely Lorentzian Quintessential Inflation (LQI) and $\alpha$-attractors in the context of Quintessential Inflation ($\alpha$-QI). Once these superheavy  particles 
have been created, they must decay into lighter ones to form a relativistic plasma, whose energy density will eventually dominate the one of the inflaton field in order to reheat after inflation our universe
with a  very high temperature, in both cases greater than $10^7$ GeV,  contrary to the usual belief that heavy masses suppress the particle production and, thus,  lead  to an inefficient reheating temperature. Finally, we will show that the over-production of Gravitational Waves (GWs) during this phase transition, when one deals with our models, 
does not disturb the Big Bang Nucleosynthesis (BBN) success.
\end{abstract}

\vspace{0.5cm}

\pacs{04.20.-q, 98.80.Jk, 98.80.Bp}
\keywords{Gravitational Particle production; Quintessential Inflation; Reheating; Gravitational Waves; Numerical Calculations.}
\maketitle
\section{Introduction}

Today, the inflationary paradigm 
is the most accepted implementation to the Big Bang (BB) theory in order
to solve a number of shortcomings associated with the standard BB cosmology, such as the horizon, the  flatness or the primordial monopole problems \cite{guth, linde}, and to explain correctly the early universe at background level.  More remarkable is the fact that inflation is also able
 to explain the  origin of inhomogeneities  in the universe as quantum fluctuations \cite{chibisov, starobinsky, pi, bardeen, Linde:1982uu}, leading to theoretical results that 
 match greatly with the  recent observational data provided by the Planck's team \cite{Planck}. 
 
 \
 
 Once having a viable theory explaining  the early universe, one can extend it in order to deal with the whole evolution of the universe, and thus, unifying its early- and late-time accelerated expansions (see \cite{odintsov, li, Copeland:2006wr} for a review of the current dark energy models). One of the most attractive scenarios able to do it is the so-called {\it Quintessential Inflation} (QI), introduced for the first time 
 by Peebles and Vilenkin  in his seminal paper \cite{pv} (see \cite{hap19} for a review of the Peebles-Vilenkin model), where  
 the idea behind  their proposal comes through the introduction of a single scalar field, also named inflaton, 
  that at early times is the responsible for inflation while at late times it allows the current cosmic acceleration via quintessence.
  %(see  \cite{Copeland:2006wr} for a review of some dark energy models).

\

Due to the simplicity of this proposal and since the behavior of the slow-roll regime is the one of an attractor -the dynamics of the model are simply obtained with the initial value of the scalar field and its derivative  at some moment during this regime-, the models of QI \cite{deHaro:2016hpl,deHaro:2016hsh,deHaro:2016ftq} -which generally only depend on two parameters-  caught  the attention of  some researchers
 who wanted  to confront QI with the observational data \cite{A,B,C,D,E,F,G,H,I,J,K,K1,K2,K3,K4,K5,K6,K7,K8}, thus becoming  a popular topic in some reduced circles.

\

In this way, dealing with QI, 
it is well-known that
all the scenarios containing a period of inflation need a reheating mechanism to match with the hot BB universe
\cite{guth} because the particles existing before the beginning of this period were completely diluted at the end of inflation resulting in a very cold universe. 
Here,  we will choose as a reheating mechanism the so-called {\it gravitational particle production}
  \cite{Parker,fmm,glm,gmm,ford,Zeldovich, Damour, Giovannini} of superheavy particles conformally coupled with gravity, which was applied to standard inflation (potentials with a deep well) in \cite{kolb, kolb1,Birrell1, hashiba, hashiba1}. However, the gravitational reheating in QI is normally applied to very light fields 
   \cite{Spokoiny, pv, A, vardayan} and only in few papers, which deal with toy discontinuous models as the Peebles-Vilenkin one, it is applied to massive particles
   \cite{H, ha, hap18, F, J} (see also \cite{Hashiba} where the authors also deals with the gravitational production of dark matter). In fact, regarding smooth QI potentials, particle creation and in particular  gravitational particle production is sometimes associated to an overbarrier problem (see for instance Section VII of \cite{kofman}), which has to be analytically studied using the complex WKB approximation, that is, the Stokes phenomenon \cite{hashiba}, whose
   real application is limited to the creation of particles by parabolic potentials \cite{kofman}.

\

 For this reason,
in the present  work we deal numerically with smooth QI potentials such as the ones provided by  Lorentzian Quintessential Inflation (LQI) \cite{benisty,benisty1,benisty2} and $\alpha$-attractors in Quintessential Inflation 
($\alpha$-QI) \cite{benisty3}. Then,  in order to get the reheating temperature,  we will calculate numerically the energy density of the superheavy particles produced during the phase transition from the end of inflation to the beginning of kination (the period where all the energy of the field becomes kinetic \cite{Joyce}). To do it, 
%The heavy massive particles due to this pre-heating will start decaying in lighter ones to form a thermal relativistic plasma. 
we will  use the well-known Hamiltonian diagonalization method (see \cite{gmmbook} for a review), showing that the time dependent 
$\beta$-Bogoliubov coefficient encodes the polarization effects associated to the creation and annihilation of the so-called {\it quasiparticles} \cite{gmmbook} and also the real particles created during the phase transition. However, through a toy model inspired in the Peebles-Vilenkin one, and presenting a discontinuity of the first derivative of the potential at the beginning of kination, we will show that
these polarization effects disappear when the universe evolves adiabatically, which happens soon after the beginning of kination. Thus, in order to calculate the energy density of the produced particles (the real particles), one can safely use the value of the $\beta$-Bogoliubov coefficient after the beginning of kination. This is the key point of our investigation, and 
we have numerically checked that this also happens for our QI smooth potentials. 

\

Another important point is the overproduction of Gravitational Waves (GWs) in QI, which are also produced during the phase transition from the end of inflation to the beginning of kination. This overproduction  in many QI models may  disturb the success of the  Big Bang  Nucleosynthesis (BBN), but, as we will show,  the reheating via gravitational particle production of superheavy particles in the QI scenarios studied in this work
prevents the incompatibilities of the BBN with the overproduction of GWs, and the reason why this happens is the fact that the gravitational production of superheavy particles, contrary to the standard belief that in analytic calculations only ultraviolet modes are taken into account,  is very efficient for long wavelength  modes, leading to a high reheating temperature able to overcome all the constraints ensuring the BBN success.

\

Finally, a few words about the viable values of the reheating temperature are in order. A lower bound for the reheating temperature comes from the fact that the
radiation-dominated  era occurs before the BBN epoch, which takes place in the  $1$ MeV  regime \cite{gkr}, and thus, the reheating temperature should naturally be greater than $1$ MeV.  On the contrary, the upper bound of this  temperature is 
dependent on the theory we are concerned with. In fact,  in some supergravity  theories  such as $\alpha$-attractors containing particles with only gravitational interactions, the late time decay of these relics  may jeopardize the success of the standard BBN \cite{lindley}. To solve this problem one has to consider 
 sufficiently low reheating temperature (of the order of $10^9$ GeV or less) \cite{eln}.
 
\

The units used in the manuscript are
 $\hbar=c=1$ and the reduced Planck's mass is denoted by $M_{pl}\equiv \frac{1}{\sqrt{8\pi G}}\cong 2.44\times 10^{18}$ GeV.

\section{The diagonalization method}
\label{sec-diagonalization}

This short section is a review of our previous work  \cite{hpa} (see also the pioneering works \cite{fmm,glm,gmm,zs} for a more detailed vision of the topic). The idea of the method goes as follows:
given a  quantum scalar field  of superheavy  particles conformally coupled to gravity, 
namely $\chi$, the Klein-Gordon (K-G) equation in the Fourier space,  which is satisfied by the modes in the flat Friedmann-Lema{\^\i}tre-Robertson-Walker (FLRW) spacetime, is given by \cite{Birrell}
\begin{eqnarray}\label{kg1}
\chi_{ k}''(\tau)+\omega^2_k(\tau) \chi_{ k}(\tau)=0,
\end{eqnarray}
%\begin{eqnarray}\label{kg0}
%\chi''+2{\mathcal H}\chi'- \nabla^2\chi+\left(m_{\chi}^2a^2+\frac{a''}{a}    \right)\chi=0,
%\end{eqnarray}
where the prime denotes the derivative with respect to the conformal time $\tau$ and $\omega_k(\tau)=\sqrt{k^2+m_{\chi}^2a^2(\tau)}$ is the time-dependent frequency being 
$m_{\chi}$  the mass of the quantum field $\chi$. 
%${\mathcal H} \equiv a'/a$, is the conformal Hubble parameter and $m_{\chi}$ is the mass of the quantum field $\chi$. 
%Now, writing the quantum field in Fourier space,
%\begin{eqnarray}
%\chi({\bf x},\tau)=\frac{1}{(2\pi)^{3/2}a}\int d^3k\left( \hat{a}_{\bf{k}}\chi_{ k}(\tau) %e^{-i\bf{k}.\bf{x}}+ 
% \hat{a}_{\bf{k}}^{\dagger}\chi_{ k}^*(\tau) e^{i\bf{k}.\bf{x}} \right),
%\end{eqnarray}
 %where $d^3 k=dk_1dk_2dk_3$, ${\bf k}=(k_1,k_2,k_3)$, ${\bf x}=(x_1,x_2,x_3)$,   $k=\sqrt{k_1^2+k^2_2+k_3^2}$ and $\hat{a}_{\bf k}$ is the annihilation operator corresponding to the 

 \
 
 As usual, the modes that define the
 vacuum state at a given initial time $\tau_i$ are the ones that minimize the energy density, so they  must
 satisfy  the conditions
 \begin{eqnarray}
 \chi_{k}(\tau_i)=
 \frac{1}{\sqrt{2\omega_k(\tau_i)}}e^{-i\int^{\tau_i} \omega_k(\bar\eta)d\bar\eta}, \quad
 \chi_{ k}'(\tau_i)=
-i \omega_k(\tau_i)\chi_{ k}(\tau_i), \end{eqnarray}
% where the time-dependent frequency is  $\omega_k(\tau)=\sqrt{k^2+m_{\chi}^2a^2(\tau)}$, and are solutions of 
%the Klein-Gordon equation
%\begin{eqnarray}\label{kg1}
%\chi_{ k}''(\tau)+\omega^2_k(\tau) \chi_{ k}(\tau)=0.
%\end{eqnarray}
and thus, 
 the vacuum expectation value of the energy density  will be  given by \cite{Bunch}
\begin{eqnarray}\label{vacuum-energy}
\langle\rho(\tau)\rangle\equiv \langle 0| \hat{\rho}(\tau)|0 \rangle=
\frac{1}{4\pi^2a^4(\tau)}\int_0^{\infty} k^2dk \left(   |\chi_{ k}'(\tau)|^2+ \omega^2_k(\tau) |\chi_{ k}(\tau)|^2-  \omega_k(\tau)        \right),
\end{eqnarray}
where, in order to obtain a finite energy density \cite{gmmbook}, we have subtracted the energy density of the zero-point oscillations of the vacuum 
$\frac{1}{(2\pi)^3a^4(\tau)}\int d^3k  \frac{1}{2} \omega_k(\tau)$.

\

Following the method developed in \cite{zs} (see also Section $9.2$ of \cite{gmmbook}),  we will write
the modes as follows,
\begin{eqnarray}\label{zs}
\chi_{k}(\tau)= \alpha_k(\tau)\frac{e^{-i\int^{\tau} \omega_k(\bar\tau)d\bar\tau}}{\sqrt{2\omega_k(\tau)}}+
\beta_k(\tau)\frac{e^{i\int^{\tau} \omega_k(\bar\tau)d\bar\tau}}{\sqrt{2\omega_k(\tau)}},\end{eqnarray}
where $\alpha_k(\tau)$ and $\beta_k(\tau)$ are the time-dependent Bogoliubov coefficients.
Now, imposing that the modes satisfy   the condition
\begin{eqnarray}
\chi_{k}'(\tau)= -i\omega_k(\tau)\left(\alpha_k(\tau)\frac{e^{-i\int^{\tau} \omega_k(\bar\tau)d\bar\tau}}{\sqrt{2\omega_k(\tau)}}-
\beta_k(\tau)\frac{e^{i\int^{\tau} \omega_k(\bar\tau)d\bar\tau}}{\sqrt{2\omega_k(\tau)}}\right),\end{eqnarray}
one can show that   the Bogoliubov coefficients must satisfy the system 
\begin{eqnarray}\label{Bogoliubovequation}
\left\{ \begin{array}{ccc}
\alpha_k'(\tau) &=& \frac{\omega_k'(\tau)}{2\omega_k(\tau)}e^{2i\int^{\tau} \omega_k(\bar\tau)d\bar\tau}\beta_k(\tau)\\
\beta_k'(\tau) &=& \frac{\omega_k'(\tau)}{2\omega_k(\tau)}e^{-2i\int^{\tau}\omega_k(\bar\tau)d\bar\tau}\alpha_k(\tau),\end{array}\right.
\end{eqnarray}
in order for the  expression (\ref{zs}) to be a solution of the equation (\ref{kg1}).

%\begin{remark}
%Since the Wronskian is conserved and $W[\chi_k(\tau_i),\chi^*_k(\tau_i)]\equiv \chi_k(\tau_i)(\chi^*_k)'(\tau_i)
%-\chi_k'(\tau_i)\chi^*_k(\tau_i)
%=i$, one can see that the Bogoliubov coefficients satisfy the equation $|\alpha_k(\tau)|^2- |\beta_k(\tau)|^2=1$.
%\end{remark}

\

Finally, inserting (\ref{zs}) into the expression for the vacuum energy density (\ref{vacuum-energy}), 
and taking into account that the Bogoliubov coefficients satisfy the equation $|\alpha_k(\tau)|^2- |\beta_k(\tau)|^2=1$,
one finds that
\begin{eqnarray}\label{vacuum-energy1}
\langle\rho(\tau)\rangle= \frac{1}{2\pi^2a^4(\tau)}\int_0^{\infty} k^2\omega_k(\tau)|\beta_k(\tau)|^2 dk,
\end{eqnarray}
where it is important to notice that $|\beta_k(\tau)|^2$ encodes the vacuum polarization effects and also the production of particles, which only happens when the adiabatic evolution breaks.  In fact, the quantity 
\begin{eqnarray}\langle N(\tau)\rangle=\frac{1}{2\pi^2 a^3(\tau)}\int_0^{\infty}k^2 |\beta_k(\tau)|^2 dk\end{eqnarray}
was named, in the Russian literature,  as the number density of {\it quasiparticles} \cite{gmmbook}, which, as we will see in next section dealing with a toy model, is very different from the number density of the produced particles because it also contains the vacuum polarization effects, that is, the creation and annihilation of pairs.

%Coming back to the equation  (\ref{Bogoliubovequation}), 
%in the first approximation taking $\alpha_k(\tau)=1$, 
%we get
%\begin{eqnarray}
%\beta_k(\tau)=\int^{\tau}\frac{\omega_k'(\eta)}{2\omega_k(\eta)}e^{-2i\int^{\eta} \omega_k(\bar\eta)d\bar\eta}d\eta.
%\end{eqnarray}

%Finally, it is important to stress   that the classical Friedmann equation is modified by the following semi-classical equation $H^2=\frac{1}{3M_{pl}^2}\left({\rho+\langle\rho\rangle}
%\right)$.

\section{Particle creation of superheavy  particles conformally coupled to gravity}
\label{sec-particle-creation}

This section is devoted to the numerical calculation of the energy density  of the gravitationally produced particles in two different QI scenarios with smooth potentials, namely:

\begin{enumerate} 
\item Lorentzian Quintessential Inflation.-

Based on the well-known Lorentzian distribution,  the authors of \cite{benisty, benisty1} considered following the ansatz 
\begin{eqnarray}\label{ansatz}
\epsilon(N)=\frac{\xi}{\pi}\frac{\Gamma/2}{N^2+\Gamma^2/4},
\end{eqnarray}
where $\epsilon$ is the main slow-roll parameter,  $N$ denotes the number of e-folds, $\xi$ is the amplitude of the Lorentzian distribution and $\Gamma$ is its width. From this ansatz, one can find the exact corresponding potential of the scalar field,   namely
\begin{eqnarray}\label{original}
\hspace{-0.5cm}V(\varphi)=\lambda M_{pl}^4\exp\left[-\frac{2\xi}{\pi}\arctan\left(\sinh
\left(\gamma\varphi/M_{pl} \right)  \right)\right]\boldsymbol{\cdot} 
\left(1-\frac{2\gamma^2\xi^2}{3\pi^3}\frac{1}{\cosh
\left(\gamma\varphi/M_{pl} \right) } \right),
\end{eqnarray}
where $\lambda$ is a dimensionless parameter and the parameter $\gamma$ is defined by
$$\gamma\equiv \sqrt{\frac{\pi}{\Gamma \xi}}.$$

Here, to simplify the structure of the potential although without modifying its properties, we set $\xi=\gamma$ and disregard the term $\left(1-\frac{2\gamma^2\xi^2}{3\pi^3}\frac{1}{\cosh
\left(\gamma\varphi/M_{pl} \right) } \right)$, which has no influence in the dynamics. Hence, we obtain the simplified version
%\begin{eqnarray}\label{PV}
%V(\varphi)=\left\{\begin{array}{ccc}
%\frac{1}{2}m^2\left(\varphi^2-M_{pl}^2+M^2\right)& \mbox{for}& \varphi\leq -M_{pl}\\
%\frac{1}{2}m^2\frac{M^6}{(\varphi+M_{pl})^4+M^4}& \mbox{for}& \varphi\geq -M_{pl},
%\end{array} \right.
%\end{eqnarray}
\begin{eqnarray}\label{LQI}
V(\varphi)=\lambda M_{pl}^4\exp\left[-\frac{2\gamma}{\pi}\arctan\left(\sinh\left(\gamma\varphi/M_{pl} \right)  \right)\right],
\end{eqnarray}
plotted in Figure \ref{fig:LQI}, and  where in order to match with the current observational data one has to choose $\lambda\sim 10^{-69}$ and $\gamma\cong 122$ (see for details \cite{benisty2}).

\

Summing up, in this scenario kination starts when $H\sim H_{kin}\cong 4\times 10^{-8} M_{pl}$ with $\varphi_{kin}\cong-0.03 M_{pl}$, inflation ends when $\varphi_{END}\cong -0.078M_{pl}$ and 
the pivot scale leaves the Hubble radius when $\varphi_*\cong -0.154 M_{pl}$. Finally, at very late times the effective Equation of State (EoS) parameter goes to $-1$, which leads to an eternal acceleration.

\begin{figure}[ht]
\includegraphics[width=0.4\textwidth]{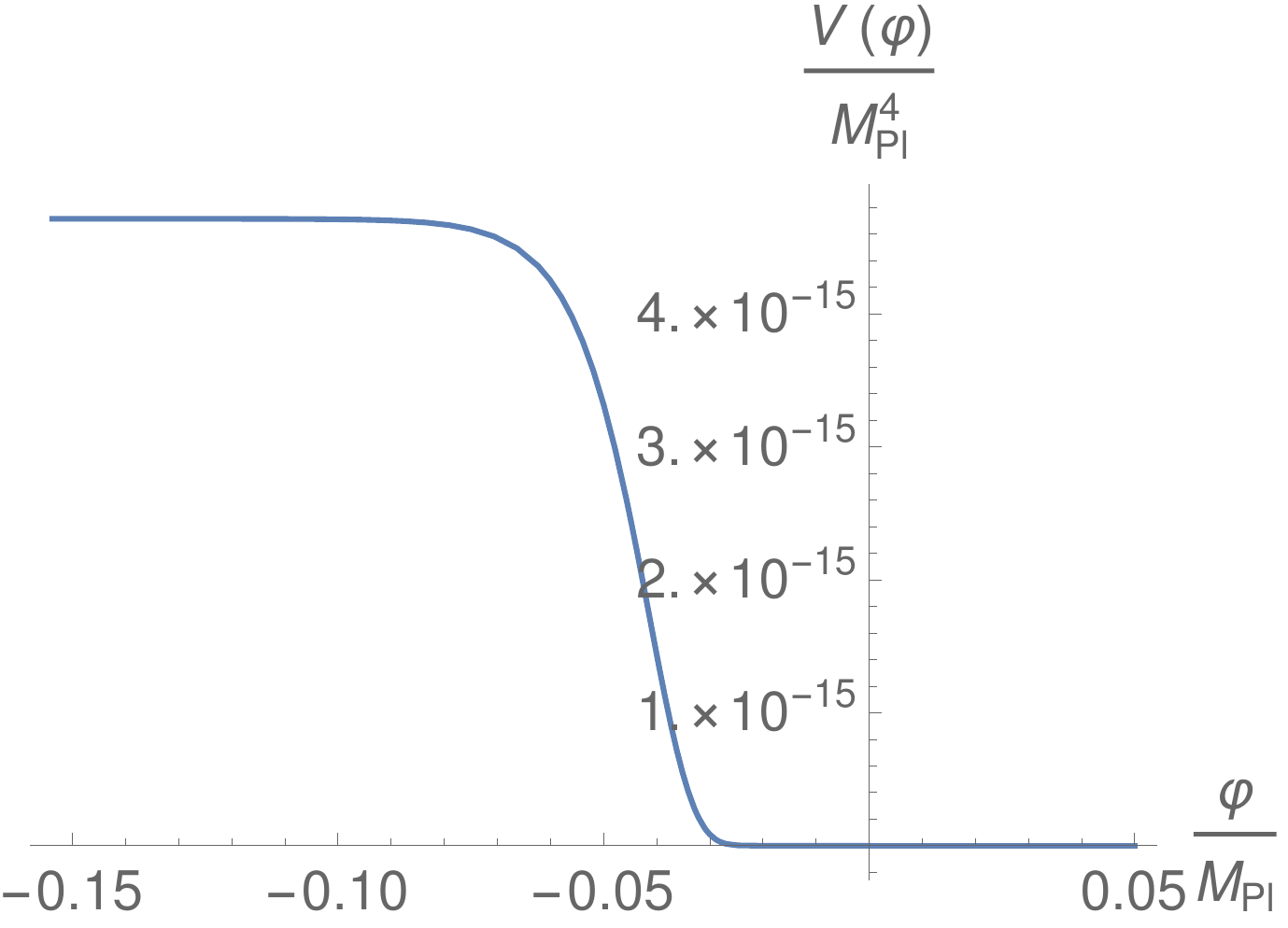}
\caption{{Plot of the Lorentzian Quintessential Inflation potential.}}
\label{fig:LQI}
\end{figure}

\item Exponential $\alpha$-attractor in Quintessential Inflation.-
%\begin{eqnarray}\label{PV2}
%V(\varphi)=\left\{\begin{array}{ccc}
%\%frac{1}{2}m^2(\varphi^2+M^2)& \mbox{for}& \varphi\leq 0\\
%\frac{1}{2}m^2\frac{M^6}{\varphi^4+M^4} &\mbox{for}& \varphi\geq 0.
%\end{array}\right.
%\end{eqnarray}

The corresponding potential is obtained, combined with  a standard   exponential potential,  from the following Lagrangian motivated by supergravity and corresponding to a non-trivial K\"ahler manifold (see for instance \cite{K} and the  references therein),
\begin{eqnarray}\label{lagrangian}
\mathcal{L}=\frac{1}{2}\frac{\dot{\phi}^2}{(1-\frac{\phi^2}{6\alpha}  )^2}M_{pl}^2-\lambda M_{pl}^4 e^{-\kappa \phi},
\end{eqnarray}
where $\phi$ is a dimensionless scalar field, and $\kappa$ and $\lambda$ are positive dimensionless constants.

\

In order that the kinetic term has the canonical  form, one can redefine the scalar field as follows,
\begin{eqnarray}
\phi= \sqrt{6\alpha}\tanh\left(\frac{\varphi}{\sqrt{6\alpha}M_{pl}}  \right),
\end{eqnarray}
obtaining the following potential plotted in Figure \ref{fig:attr},
\begin{eqnarray}\label{alpha}
V(\varphi)=\lambda M_{pl}^4e^{-n\tanh\left(\frac{\varphi}{\sqrt{6\alpha}M_{pl}} \right)},
\end{eqnarray}
where we have introduced the notation $n \equiv\kappa \sqrt{6\alpha}$, and  by taking  $\alpha\sim 10^{-2}$ one has to choose $n\sim 10^2$ and $\lambda \sim 10^{-66}$ in order to match with the observational data (see  \cite{benisty3} for details).

\

Finally, 
it is interesting to note that for this case kination starts later than in LQI, more precisely, when $\varphi_{kin}\cong -0.5 M_{pl}$ with $H_{kin}\sim 4\times 10^{-7} M_{pl}$. However, the end of inflation and the horizon crossing occurs earlier than in LQI,
$\varphi_{END}\cong -0.89 M_{pl}$ and $\varphi_*\cong -1.7 M_{pl}$ respectively. 

%\begin{figure}[ht]
%\includegraphics[width=0.4\textwidth]{pot2}
%\includegraphics[width=0.4\textwidth]{pot22}
%\caption{{\color{red}Plot of the Peeble-Vilenkin potentials in \eqref{PV}, in two different scales, using the values derived below.}}
%\label{fig:PV1}
%\end{figure}

\begin{figure}[ht]
\includegraphics[width=0.4\textwidth]{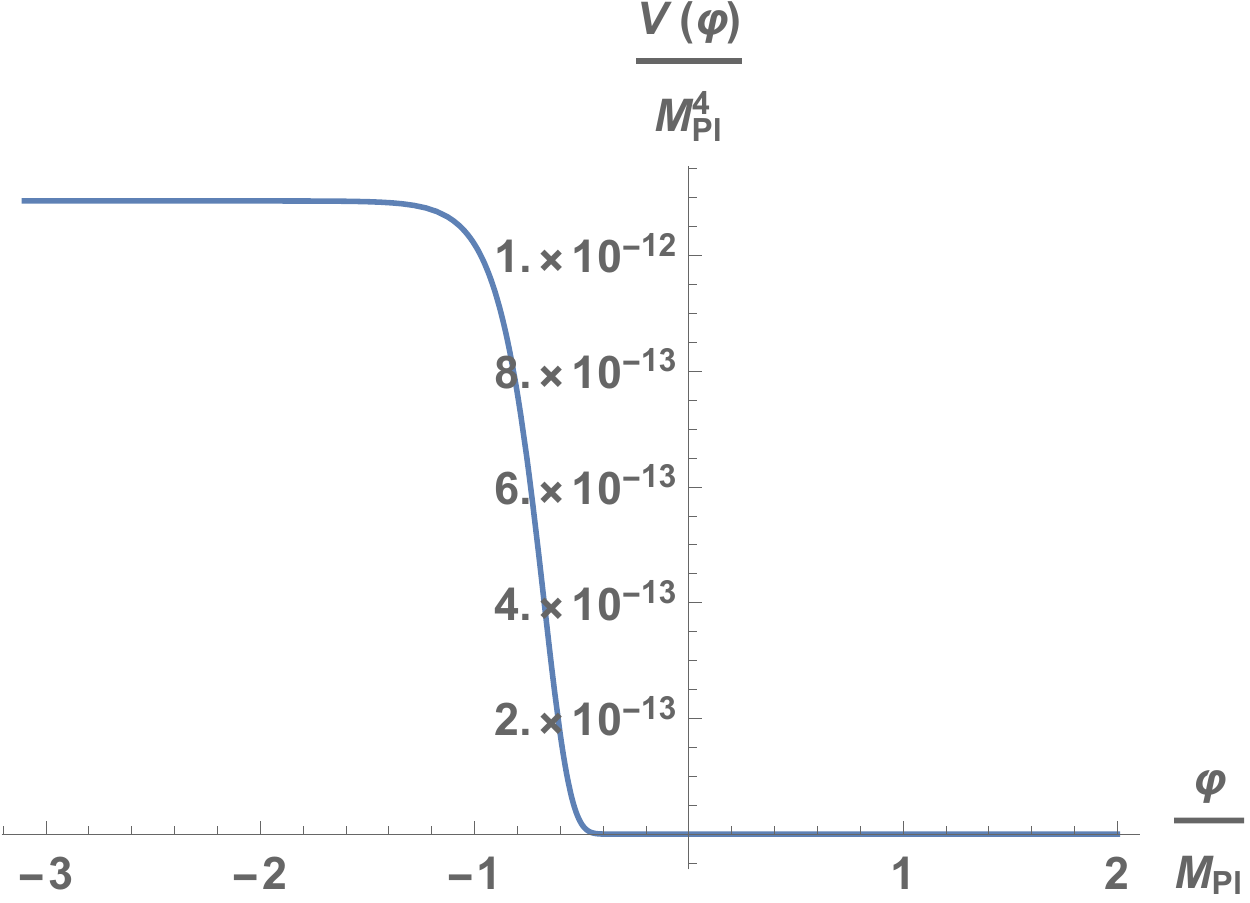}
\caption{{Plot of the Exponential $\alpha$-attractor potential.}}
\label{fig:attr}
\end{figure}

\end{enumerate}

\subsection{A toy model}

Before dealing with our models and in order to
understand better the gravitational particle production, 
we warm up revewing   a toy potential inspired in the Peebles-Vilenkin model \cite{hpa},
{\begin{eqnarray} \label{toy}
V(\varphi)=\left\{\begin{array}{ccc}
\frac{1}{2}m^2 \left( \varphi^2-M_{pl}^2+ {M}^2\right) & \mbox{for} & \varphi\leq -M_{pl}\\
\frac{1}{2}m^2 \frac{M^6}{(\varphi+M_{pl})^{4}+M^4} &\mbox{for} & \varphi\geq -M_{pl},\end{array}
\right.
\end{eqnarray}
where $m\sim 5\times 10^{-6} M_{pl}$ is the mass of the inflaton field and $M\cong 20$ GeV  is a very small mass needed to match the theoretical results provided  by the model with the current cosmic acceleration
\cite{hap19}.

}

\

The toy model contains a discontinuity of the first derivative of the potential  at the beginning of kination, i.e., when $\varphi=0$. Then, according to  the conservation equation,  the second temporal derivative of the scalar field is discontinuous at the beginning of kination, as well as the second temporal derivative of the Hubble parameter.  Consequently,  the third derivative of the frequency $\omega_k(\tau)$ (which depends on the scale factor) is discontinuous at the beginning of kination, namely $\tau_{kin}$, which is the  moment when particles are gravitationally  created because it is when the adiabatic evolution is broken.

\
  
On the other hand, before  performing analytic calculations with this toy model, the following remark, which helps us to understand the more  realistic ones,  is needed:  we have to make some assumptions  to be sure that the polarization effects do not affect the evolution of the inflaton field during the slow-roll
period, that is, during the slow-roll regime we will demand that  the polarization effects will be sub-dominant and do not affect the dynamics of the inflaton field.  Analytically,  a condition that ensures that polarization effects are sub-dominant is  $H/m_{\chi}\ll 1$,  which implies that $\omega_k'/\omega_k^2\ll 1$, i.e., an adiabatic evolution, and where once again $m_{\chi}$ is the mass of superheavy field $\chi$. Taking into account that a classical picture of the universe (quantum gravitational effects could be neglected)  appears at GUT scales with $H_{GUT}\cong 5\times 10^{-5} M_{pl}\cong 10^{14}$ GeV, we have to choose superheavy massive fields with  a mass greater or equal than  $m_{\chi}\sim 5\times 10^{-4} M_{pl}\cong 10^{15}$ GeV,
 which is a mass of the same order as those of the vector mesons responsible for transforming quarks into leptons in simple theories with SU(5) symmetry \cite{lindebook}.

\

Therefore, once we have chosen the mass  of the $\chi$-field, 
in order to obtain the value of the $\beta$-Bogoliubov coefficient
we come  back to the equation  (\ref{Bogoliubovequation}) and, 
in the first approximation,  we take  $\alpha_k(\tau)=1$, 
getting
\begin{eqnarray}
\beta_k(\tau)=\int^{\tau}\frac{\omega_k'(\eta)}{2\omega_k(\eta)}e^{-2i\int^{\eta} \omega_k(\bar\eta)d\bar\eta}d\eta,
\end{eqnarray}
which, after integration by parts, yields before the beginning of kination
\begin{eqnarray}
\beta_k(\tau)
= \left(-\frac{\omega'_k(\tau)}{4i\omega_k^2(\tau)}+\frac{1}{8\omega_k(\tau)}\left(\frac{\omega'_k(\tau)}{\omega_k^2(\tau)}\right)'
+\frac{1}{16i\omega_k(\tau)}\left(\frac{1}{\omega_k(\tau)}\left(\frac{\omega'_k(\tau)}{\omega_k^2(\tau)}\right)'\right)'+....
  \right)e^{-2i\int^{\tau} \omega_k(\bar\eta)d\bar\eta}.
  \end{eqnarray}

\

However, after the beginning of  kination the $\beta$-Bogoliubov coefficient must be given by
\begin{eqnarray}\label{Bogoliubov}
\beta_k(\tau)=
\left(-\frac{\omega'_k(\tau)}{4i\omega_k^2(\tau)}+\frac{1}{8\omega_k(\tau)}\left(\frac{\omega'_k(\tau)}{\omega_k^2(\tau)}\right)'
+\frac{1}{16i\omega_k(\tau)}\left(\frac{1}{\omega_k(\tau)}\left(\frac{\omega'_k(\tau)}{\omega_k^2(\tau)}\right)'\right)'+....
  \right)e^{-2i\int^{\tau} \omega_k(\bar\eta)d\bar\eta}+C,\end{eqnarray}
where the constant $C$ has to be chosen in order that the $\beta$-Bogoliubov coefficient becomes continuous at $\tau_{kin}$ because the equation (\ref{Bogoliubovequation})
is a first order differential equation, and mathematically it is necessary  to demand the solution to be continuous.
Thus, after some cumbersome calculations (see \cite{hpa} for details) one has
\begin{eqnarray}\label{constant}
C= \left(
\frac{m_{\chi}^2m^3a^5_{kin}}{16i\omega^5_k(\tau_{kin})}+
....
  \right)e^{-2i\int^{\tau_{kin}} \omega_k(\bar\eta)d\bar\eta},
 \end{eqnarray}
where we have introduced the definition $a_{kin}\equiv a(\tau_{kin})$.

\

The terms of the $\beta$-Bogoliubov coefficient different from $C$ lead to sub-leading geometric quantities in the energy density. Effectively, the term $-\frac{\omega'_k(\tau)}{4i\omega_k^2(\tau)}$
leads to the following contribution to the energy density, $\frac{m_{\chi}^2 H^2}{96 \pi}$, which is negligible compared with    $H^2M_{pl}^2$. The same happens with  $\frac{1}{8\omega_k(\tau)}\left(\omega'_k(\tau)/\omega_k^2(\tau)\right)'$ leading  to a term of order $H^4$,  which satisfies  ${H^4}\ll H^2M_{pl}^2$.
The product of the first and second term generates in the right-hand side of the modified semi-classical Friedmann  equation a term of the order 
${H^3m_{\chi}}$, which is also sub-leading compared with $H^2M_{pl}^2$. Finally, the third term of (\ref{Bogoliubov}) leads in the right-hand side of the semi-classical Friedmann equation to the sub-leading term $\frac{H^6}{m_{\chi}^2}$.

\

Fortunately, this does not happen with $C$, whose leading term  gives the main contribution  of the vacuum energy density due to the gravitational particle production. In fact, the time dependent terms, which as we have already shown  are always sub-leading,  are vacuum polarization effects, and they rapidly disappear in the adiabatic regime, that is, soon after the beginning of kination  
 $|\beta_k(\tau)|^2$ approaches  to $|C|^2$, obtaining
\begin{eqnarray}
\langle\rho(\tau)\rangle\cong \left\{\begin{array}{ccc}
0& \mbox{ when} & \tau< \tau_{kin} \\
10^{-5}\left(\frac{m}{m_{\chi}}  \right)^2m^4\left( \frac{a_{kin}}{a(\tau)} \right)^3 & \mbox{ when}    & \tau\geq \tau_{kin},
\end{array}\right.
\end{eqnarray}
which at the beginning of kination is sub-dominant with respect to the energy density of the inflaton but it will eventually dominate because the one of the inflaton decreases during kination as $a^{-6}(\tau)$.

\

Finally, 
in order to understand better these results it is useful to recall, as we have already explained,  that
the authors of the diagonalization method assume that, during the whole evolution of the universe, quanta named {\it quasiparticles} are created and annihilated due to the interaction with the quantum field with gravity \cite{gmmbook}, i.e., this is a vacuum polarization effect where pairs are created and annihilated. And, following this interpretation, the number density of the created {\it quasiparticles} at time $\tau$ is given by
$\langle N(\tau)\rangle=\frac{1}{2\pi^2 a^3(\tau)}\int_0^{\infty}k^2 |\beta_k(\tau)|^2 dk$. However, one has to be very careful with this interpretation and especially keep in mind that, as we have already pointed out in our toy model (\ref{toy}), 
real particles are only created when the adiabatic regime breaks. Effectively, before the beginning of kination, i.e., before the break of the adiabatic evolution, the main term of the $\beta$-Bogoliubov coefficient is given by $-\frac{\omega'_k(\tau)}{4i\omega_k^2(\tau)}$,
whose contribution to the energy density is $\frac{m_{\chi}^2 H^2}{96 \pi}$, and to the number density of the {\it quasiparticles}  $\frac{m_{\chi} H^2}{512 \pi}$, and thus,
at any time $\tau$ before the beginning of kination $\langle\rho(\tau)\rangle\not=m_{\chi}\langle N(\tau)\rangle$, meaning that the {\it quasiparticles} do not evolve as real massive particles. On the contrary, during kination the polarization effects disappear and 
the leading term of 
$\langle N(\tau)\rangle$ is given by $ 10^{-5}\left(\frac{m}{m_{\chi}}  \right)^3m^3\left( \frac{a_{kin}}{a(\tau)} \right)^3$, having  $\langle\rho(\tau)\rangle=m_{\chi}\langle N(\tau)\rangle$
and a decay of $a^{-3}(\tau)$,
which justifies the interpretation of massive particle production.

\

\subsection{ Gravitational particle creation in Lorentzian Quintessential Inflation}

Now we are ready  to calculate the particle production for our QI models.
Coming back to the LQI potential (\ref{LQI}),
first of all we have integrated numerically the conservation equation for the inflaton field, namely
\begin{eqnarray}\label{conservation}
\ddot{\varphi}+3H\dot{\varphi}+V_{\varphi}=0,
\end{eqnarray}
where $H=\frac{1}{\sqrt{3}M_{pl}}\sqrt{\frac{\dot{\varphi}^2}{2}+V(\varphi)  }$, with initial conditions at the horizon crossing,
 i.e., when the pivot scales leaves the Hubble radius. Recall that in that moment
the system is in the slow-roll phase and, since this regime is an attractor, one only has to take initial conditions in the basin of attraction of the slow-roll solution, for example,
$\varphi_*=-0.154 M_{pl}$ and $\dot{\varphi}_*=-\frac{V_{\varphi}(\varphi_*)}{3H_*}$, where the ``star" denotes that the quantities are evaluated
at the horizon crossing.
%{\color{red} crec que és millor agafar $\dot{\varphi}_*=-\frac{V_{\varphi}(\varphi_*)}{3H_*}$, i així ens asegurem que la solució està a la conca d'atracció de la slow-roll. Ho dic pel model $\alpha$-QI, per l'altre em quadra tot  }.

\

Once we  have obtained the evolution of the background, and in particular the evolution of the Hubble rate, we compute the evolution of the scale factor, which is 
given by 
\begin{eqnarray}
a(t)=a_*e^{\int_{t_*}^t H(s)ds},
\end{eqnarray}
where we have chosen as the value of the scale factor at the horizon crossing $a_*=1$.

\

\begin{figure}[ht]
\includegraphics[width=0.4\textwidth]{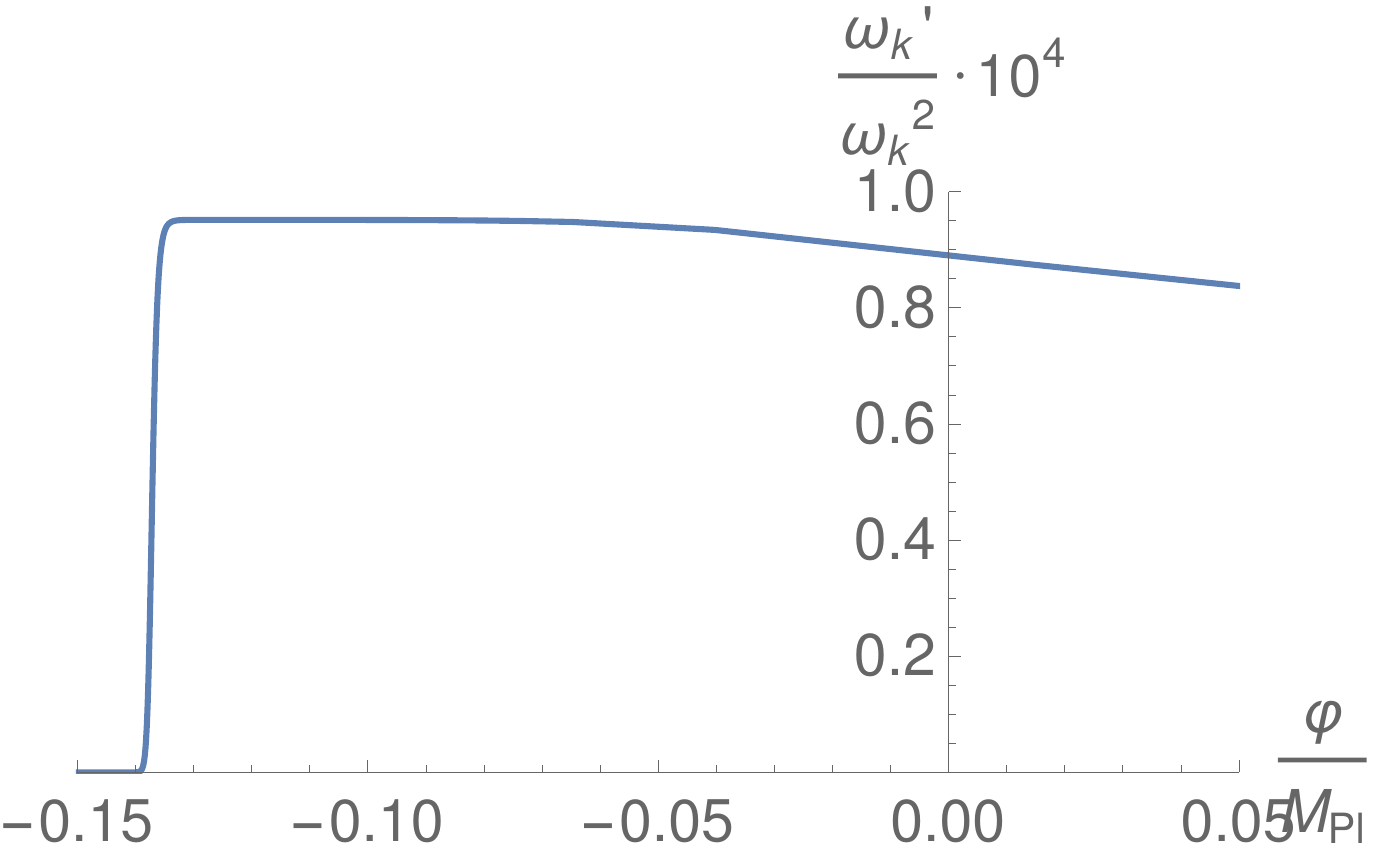}
\includegraphics[width=0.4\textwidth]{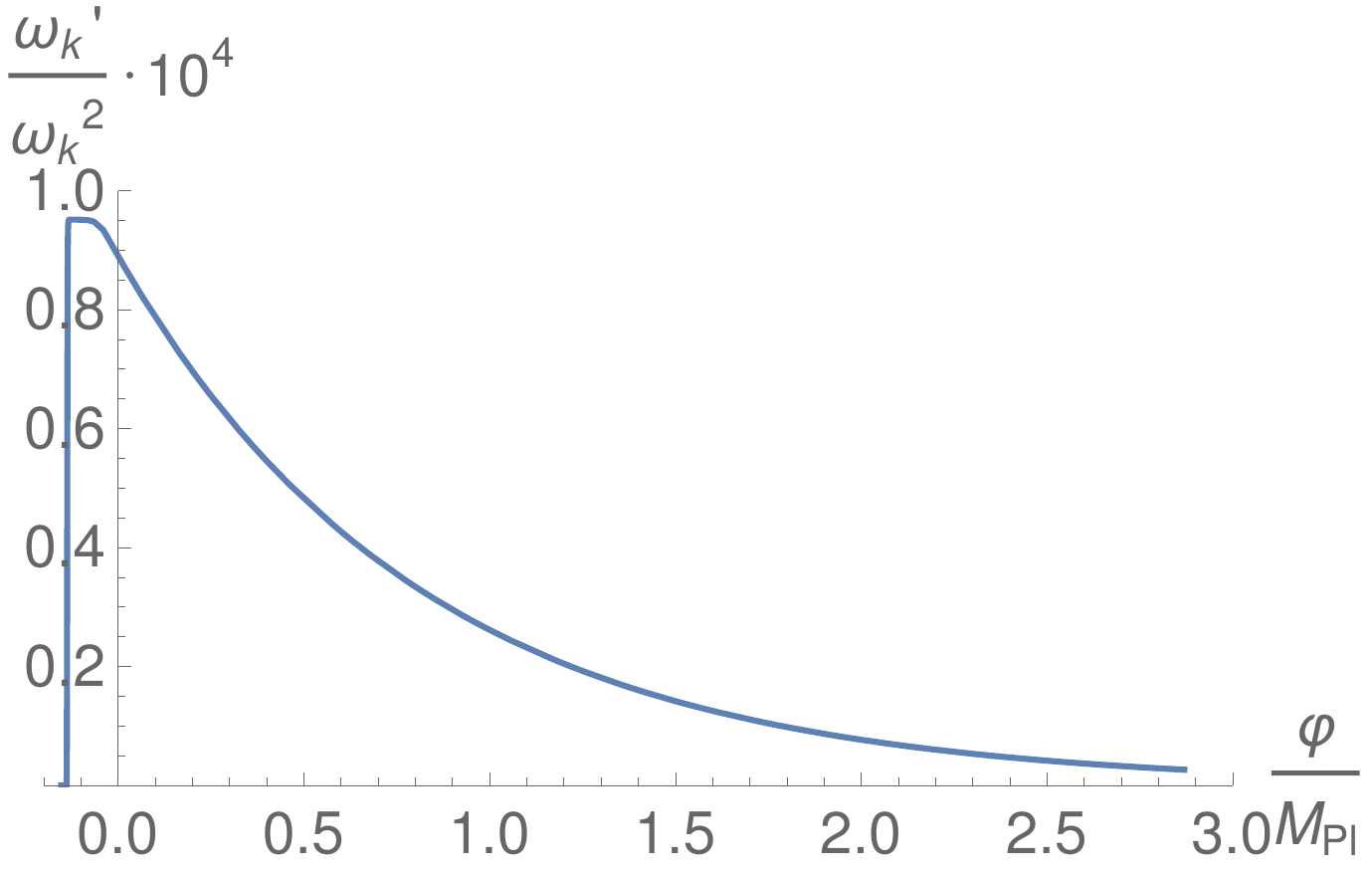}
\caption{Plot of the adiabatic evolution for a heavy field with mass $m_{\chi}\cong 10^{15}$ GeV, when the background is given by the Lorentzian Quintessential Inflation potential. The value $k=a_{kin}H_{kin}$ has been used for the quantities $a_{kin}$, $H_{kin}$ of this model. { On the left, with the same scale as the one used in Figure \ref{fig:LQI} and on the right with a wider range of values of the scalar field in order to appreciate the extent of the non-adiabatic region.
}}
\label{fig:adiabaticLQI}
\end{figure}

From the evolution of the scale factor, 
 we can see in Figure  \ref{fig:adiabaticLQI} that a spike appears in the plot of the quantity $\omega_k'/\omega_k^2$ during the phase transition from the end of inflation to the beginning of kination, that is, at that moment when the adiabatic evolution is broken and particles are gravitationally produced.

%{\color{red} Tot aixo que ve s'ha d'explicar amb molt mes detall: dir com es la $|\beta(\tau)_k|^2$, potser fer un grafic per algun valor de $k$, dir el rang de valors de la $k$ on te valors significants.....}

\

Then, from the knowledge acquired studying  our  toy model (\ref{toy}), we have numerically solved the equation (\ref{Bogoliubovequation}), 
with initial conditions  $\alpha_k(\tau_*)=1$ and $\beta_k(\tau_*)=0$ at the horizon crossing (there were neither particles nor polarization effects at that moment because during the slow-roll regime the derivatives of the Hubble rate are negligible compared with the powers of $H$, i.e.,  the system is in the adiabatic regime).

\

{ In order to get rid of complex exponentials we have transformed equation \eqref{Bogoliubovequation} into a second order differential equation, namely
\begin{eqnarray}
\left\{ \begin{array}{cc}
     &  \alpha_k''(\tau)=\alpha_k'(\tau)\left(\frac{\omega_k''(\tau)}{\omega_k'(\tau)} - \frac{\omega_k'(\tau)}{\omega_k(\tau)}+2i\omega_k(\tau) \right)+ \left(\frac{\omega_k'(\tau)}{2\omega_k(\tau)} \right)^2\alpha_k(\tau)\\
     & \beta_k''(\tau)=\beta_k'(\tau)\left(\frac{\omega_k''(\tau)}{\omega_k'(\tau)} - \frac{\omega_k'(\tau)}{\omega_k(\tau)}-2i\omega_k(\tau) \right)+ \left(\frac{\omega_k'(\tau)}{2\omega_k(\tau)} \right)^2\beta_k(\tau)
\end{array}\right..
\end{eqnarray}

Given that $\alpha_k(\tau_*)=1$ and $\beta_k(\tau_*)=0$ leads to $\alpha_k'(\tau_*)=0$, we are interested in solving the equation for $\alpha_k(\tau)$, which can be split into the real and imaginary form in the following way,
\begin{eqnarray}
\left\{ \begin{array}{cc}
     &  \alpha_{k,Re}''(\tau)=\alpha_{k,Re}'(\tau)\left(\frac{\omega_k''(\tau)}{\omega_k'(\tau)} - \frac{\omega_k'(\tau)}{\omega_k(\tau)}\right)-2\omega_k(\tau)\alpha_{k,Im}'(\tau) + \left(\frac{\omega_k'(\tau)}{2\omega_k(\tau)} \right)^2\alpha_{k,Re}(\tau)\\
     & \alpha_{k,Im}''(\tau)=\alpha_{k,Im}'(\tau)\left(\frac{\omega_k''(\tau)}{\omega_k'(\tau)} - \frac{\omega_k'(\tau)}{\omega_k(\tau)}\right)+2\omega_k(\tau)\alpha_{k,Re}'(\tau) + \left(\frac{\omega_k'(\tau)}{2\omega_k(\tau)} \right)^2\alpha_{k,Im}(\tau)
\end{array}\right.,
\end{eqnarray}
and then $|\beta_k(\tau)|^2=|\alpha_k(\tau)|^2-1$ because of the well-known conservation property of the Wronskian. For the value $k=a_{kin}H_{kin}$, we obtain in Figure \ref{fig:bogoliubov} that $|\beta_k(\tau)|^2$ stabilizes soon to a non-zero value after the beginning of kination, containing only particle production effects. We have numerically verified that this happens for the range $0.05\lesssim\frac{k}{a_{kin}H_{kin}}\lesssim7\times 10^4$, which leads to values of $|\beta_k|^2$ of the order of $10^{-10}$ and $10^{-11}$.

\begin{figure}[ht]
    \centering
    \includegraphics[width=0.5\textwidth]{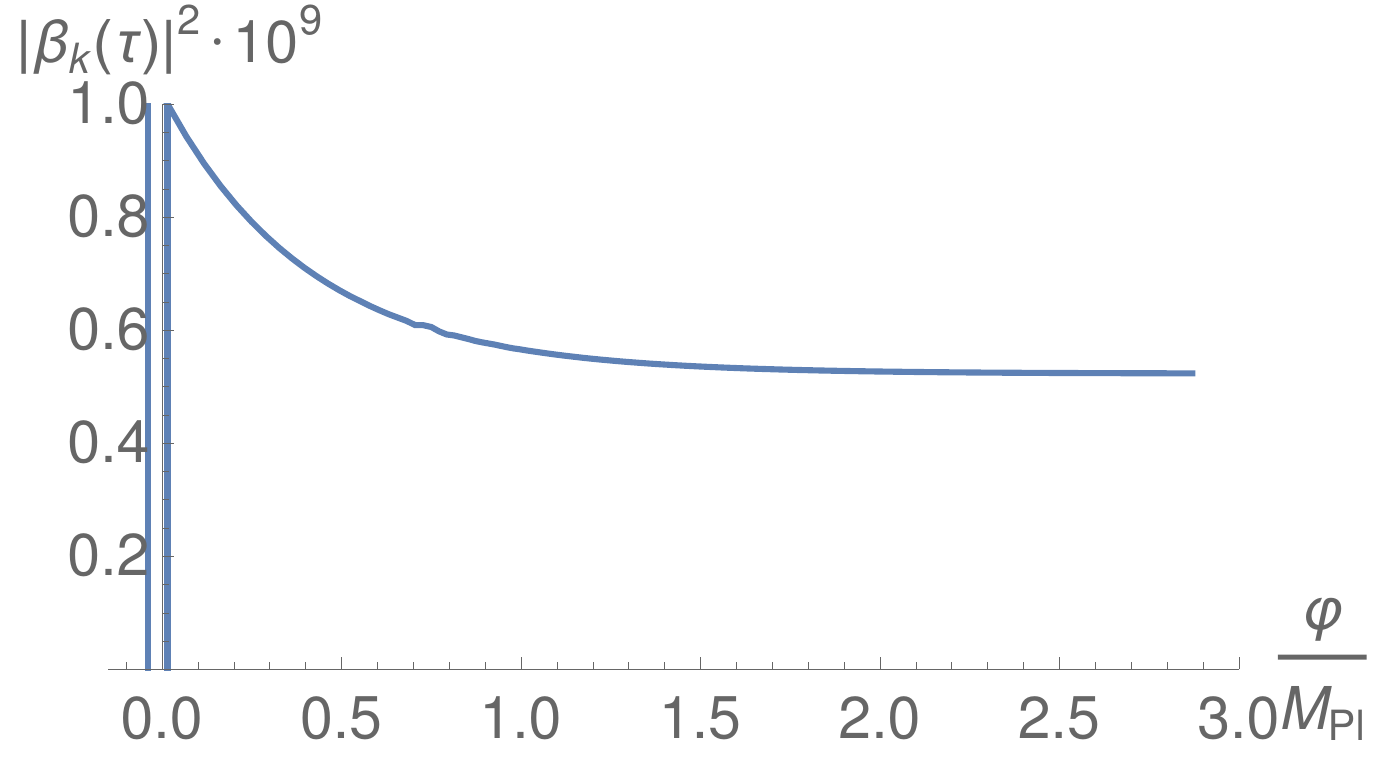}
    \caption{Evolution of $|\beta_k(\tau)|^2$}
    \label{fig:bogoliubov}
\end{figure}

}

\

Then, introducing these values of the $\beta$-Bogoliubov coefficient in 
energy density 
(equation (\ref{vacuum-energy1})),  we have obtained a vacuum energy density of the order of $10^{44} \mbox{ GeV}^4$. So, the energy density of the produced particles evolves as 
\begin{eqnarray}
\langle\rho(\tau)\rangle=\langle\bar{\rho}\rangle\left(\frac{\bar{a}}{a(\tau)}\right)^3,
\end{eqnarray}
where $\langle\bar{\rho}\rangle\cong 10^{44} \mbox{ GeV}^4\cong 3\times 10^{-30} M_{pl}^4$ and $\bar{a}$ are, respectively, the energy density of the produced particles 
and the value of the scale factor at the end of the non-adiabatic phase ($\varphi\sim  M_{pl}$, which coincides with the stabilization of the $\beta$-Bogoliubov coefficient). Finally, the energy density of the background at this moment is given by
$\bar{\rho}_{\varphi}=3\bar{H}^2M_{pl}^2\cong 2\times 10^{57}\mbox{ GeV}^4\cong 7\times 10^{-17} M_{pl}^4$, showing that the energy density of the produced particles is sub-leading close to  the beginning of kination, but will eventually be dominant because, during the kination regime, the energy density of the inflaton field decreases as $a^{-6}$.

\

\begin{remark}
Note that the values of the wavenumber $k$, that leads to significant values of the $\beta$-Bogoliubov coefficient, are
not in the ultraviolet regime. On the contrary, analytic calculations only deal with the ultraviolet spectrum leading to an insignificant value of the $\beta$-Bogoliubov coefficient for heavy masses. For this reason, without taking into account the long wavelengths,
it is usual to assume that the heavy masses suppress the particle production, thus leading to an inefficient reheating temperature. However, as we will see, this is not the case when one performs the numerical calculations considering all the spectrum of values of the wavenumber.
\end{remark}

\

\begin{figure}[ht]
\includegraphics[width=0.48\textwidth]{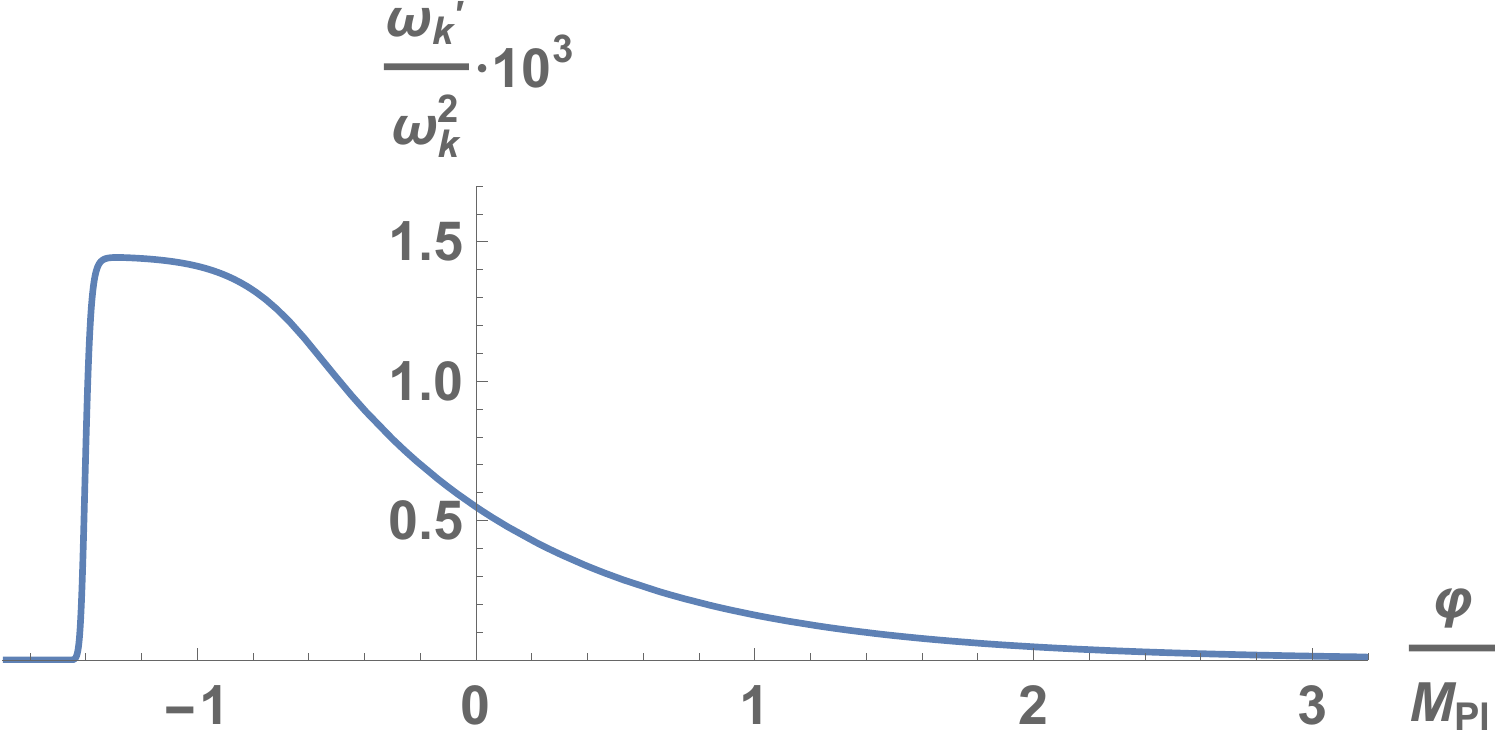}
\includegraphics[width=0.51\textwidth]{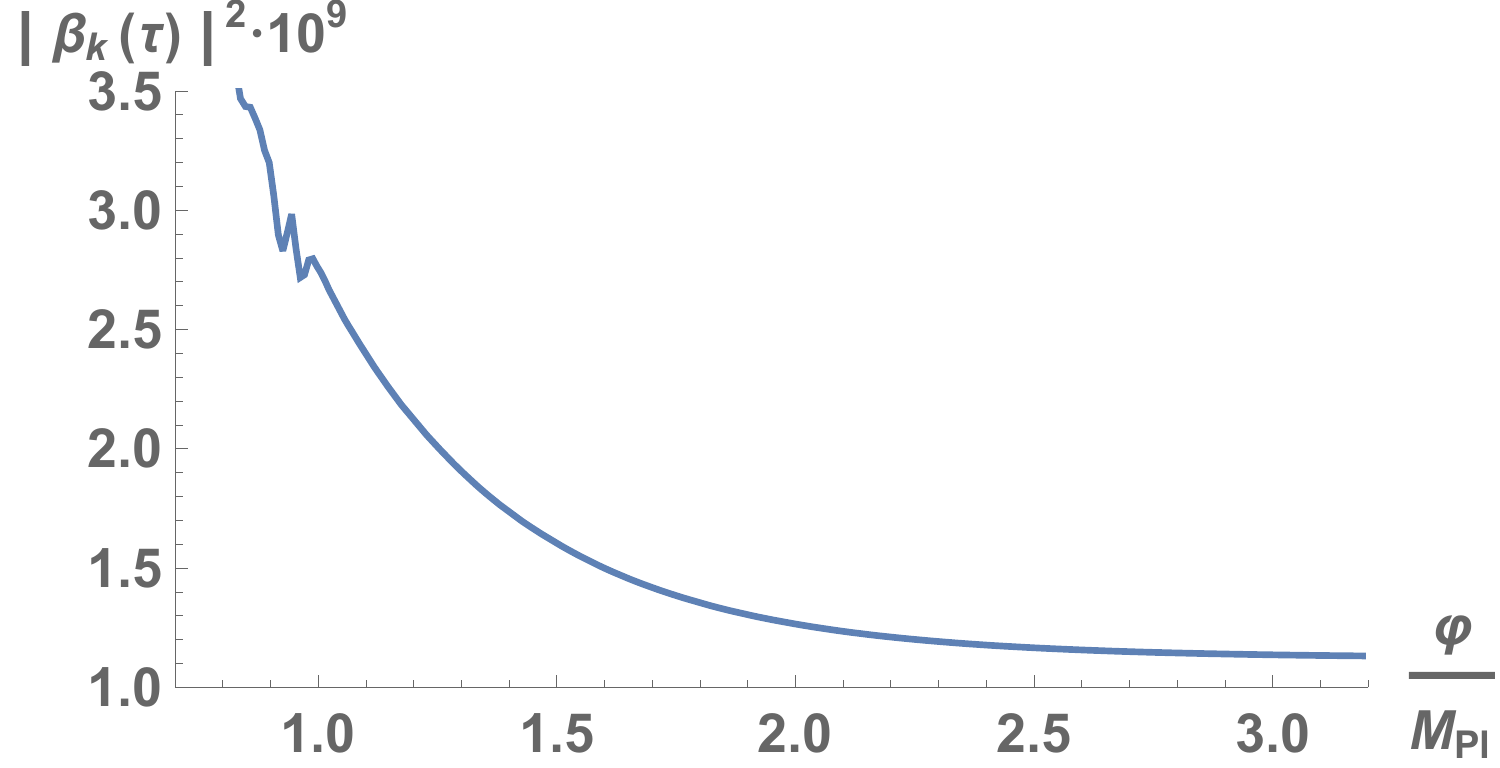}
\caption{Plot of the adiabatic evolution(left) and the $\beta$-Bogoliubov coefficient(right) for a heavy field with mass $m_{\chi}\cong 10^{15}$ GeV, when the background is given by the exponential $\alpha$-attractor potential. \\
Here we have used the value $k=a_{kin}H_{kin}$, since we have observed that particle creation takes place for modes in the range $1\lesssim\frac{k}{a_{kin}H_{kin}}\lesssim 10^3$.
%{\color{red} aquí s'hauria de fer el numèric per $k=m_{\chi}$ (amb les condicions inicial al horizon crossing que he dit abans, per assegurar que està a la conca d'atracció), i el rang de valors de $k$ l'escriuria així: $0\leq \frac{k}{m_{\chi}}\leq 1$}
%\\
%Note that the numeric turns out to be more cumbersome in this case but, however, we have been able to evolve the system long enough in order to take into account all the non-adiabatic region.
}
\label{fig:adiabaticattr}
\end{figure}

\begin{remark}
In the case of the $\alpha$-attractors in the context of Quintessential Inflation, we have numerically obtained that
%, contrary to LQI, the end of the non-adiabatic phase is a little before the beginning of kination, and 
the  corresponding energy densities are of the order
$\langle\bar{\rho}\rangle\sim 3\times 10^{-27}M_{pl}^4$ and $\bar{\rho}_{\varphi}\sim 6\times 10^{-17} M_{pl}^4$ (see Figure \ref{fig:adiabaticattr})
\end{remark}

%{\color{red} Aqui si ets capaç de fer el mateix numeric pel potencial del $\alpha$-QI, poodriem posar un remark dient el que s'obte. Que suposo ha de ser similar}

\

\

We would like to finish this subsection showing more details about the function 
$
\frac{\omega_k'}{\omega_k^2}=\frac{m_{\chi}^2a^3H}{\omega_k^3}$. We have calculated its temporal derivative, obtaining:
\begin{eqnarray}
\frac{d}{dt}\left( \frac{\omega_k'}{\omega_k^2} \right)=\frac{3H^2a^3m_{\chi}^2}{\omega_k^3}\left(
\frac{1-w_{eff}}{2}-\frac{m_{\chi}^2a^2}{\omega_k^2}\right),
\end{eqnarray}
where $w_{eff}$ denotes the effective Equation of State (EoS) paramenter, i.e, the ratio of the pressure to the energy density. We will easily see that during kination, i.e. when $w_{eff}=1$, this is a decreasing function, and is only an increasing function for the modes satisfying 
\begin{eqnarray}\label{mode}
k>\sqrt{\frac{1+w_{eff}}{1-w_{eff}}}m_{\chi}a.
\end{eqnarray}

Then, for the relevant models in LQI that contribute to the energy density of produced particles ($0.05\lesssim \frac{k}{a_{kin}H_{kin}}\lesssim 7\times 10^4$), the condition (\ref{mode}) is fulfilled  during all the inflationary regime (see Fig \ref{fig:adiabaticLQI} and recall that in LQI inflation ends when $\varphi_{END}\cong-0.078$), and thus one obtains the spike given in Fig \ref{fig:adiabaticLQI}. 

\

In addition, for the relevant modes,  at the horizon crossing the quantity 
$\frac{\omega_k'}{\omega_k^2}$ is extremely small and, since it is an increasing function during inflation, the quantity  $\frac{\omega_k'}{\omega_k^2}$
is negligible before the horizon crossing, that is, during this period there is no particle production and the polarization effects are negligible, for this reason we can ensure that before the horizon crossing the relevant modes of the field $\chi$ are in the vacuum. After the horizon crossing, polarization effects appear and, later, during the phase transition, particles are created, as we will see in Fig \ref{fig:bogoliubov}. Finally, the $\beta$-Bogoliubov coefficient stabilizes (at the same time that the non-adiabatic regime finishes (right plot of Fig \ref{fig:adiabaticLQI})) during kination, that is, the polarization effects are negligible and it only encodes the particle creation.

\

And exactly the same happens in the $\alpha$-QI scenario.

\section{The reheating process}
\label{sec-reheating}

After the production of the heavy massive particles with masses around $10^{15}$ GeV, they have to decay into lighter ones which after the thermalization process form a relativistic plasma that depicts our hot universe. Two different situations may arise, as follows: 
\begin{enumerate}
\item The decay, which occurs at time $\tau_{dec}$, is before the end of the kination regime,  when the energy density of the inflaton becomes equal to the one of the $\chi$-field.
\item The decay is after the end of the kination regime.
\end{enumerate}

%Here we consider the decay of the  $\chi$-field into fermions ($\chi\rightarrow \psi\bar\psi$), then the decay rate will be given by \cite{lindebook}
%${\Gamma}=\frac{h^2 m_{\chi}}{8\pi}$ and the decay is finished at $\tau_{dec}$ when $\Gamma\sim H(\tau_{dec})\equiv H_{dec}$.

\subsection{Decay before the end of kination}
\label{sebsec1-reheating}

Let us begin the discussion with the LQI potential. In this case,
 the energy density of the background, i.e. the one of the inflaton field,  and the one of the relativistic plasma, when the decay is finished, 
that is
 when ${\Gamma}\sim H_{dec}=\bar{H}\left(\frac{\bar{a}}{a_{dec}} \right)^3\cong
   10^{10}\left(\frac{\bar{a}}{a_{dec}}\right)^3 $ GeV $\sim 5\times 10^{-9}\left(\frac{\bar{a}}{a_{dec}}\right)^3 M_{pl}$ , will be
\begin{eqnarray}\label{LQIrho}
\rho_{\varphi, dec}=3{\Gamma}^2M_{pl}^2 \quad \mbox{and} \quad \langle\rho_{dec}\rangle\cong 3\times 10^{-30} 
\left(\frac{\bar{a}}{a_{dec}}\right)^3M_{pl}^4 \sim 10^{-21} \Gamma M_{pl}^3.
\end{eqnarray}

Imposing that the end of the decay precedes the end of kination, that means, $ \langle\rho_{dec}\rangle\leq \rho_{\varphi, dec}$, one gets
$
\Gamma\geq 10^{-21} M_{pl}.
$

\

 In addition, since the decay is after the beginning of the kination, one has $\Gamma\leq H_{kin}\cong 4\times 10^{-8} M_{pl}$. So, we have the following bound for the decay rate,
\begin{eqnarray}\label{bound}
 10^{-21} M_{pl}\leq \Gamma \leq 4\times 10^{-8} M_{pl}.\end{eqnarray}

Finally,  the reheating temperature, i.e., the temperature of the universe when the relativistic plasma in thermal equilibrium starts to dominate, 
which happens when $\rho_{\varphi, reh}=\langle \rho_{reh}\rangle$, can be calculated as follows:
Since after the decay the evolution of the respective energy densities is given by
\begin{eqnarray}
\rho_{\varphi, reh}=\rho_{\varphi, dec}\left(a_{dec}/a_{reh}\right)^6,\qquad 
\langle \rho_{reh}\rangle=\langle \rho_{dec}\rangle \left(a_{dec}/a_{reh}\right)^4,\end{eqnarray}
we will have
$\frac{\langle \rho_{dec}\rangle}{\rho_{\varphi,dec}}=\left(a_{dec}/a_{reh}\right)^2,$
and thus, the reheating temperature will be
\begin{eqnarray}\label{reheating1}
 T_{reh}=  \left(\frac{30}{\pi^2g_{reh}} \right)^{1/4}
 \langle\rho_{reh}\rangle^{\frac{1}{4}}= 
 \left(\frac{30}{\pi^2g_{reh}} \right)^{1/4}
 \langle\rho_{dec}\rangle^{\frac{1}{4}}
 \sqrt{\frac{\langle\rho_{dec}\rangle}{\rho_{\varphi,dec}}} 
\sim 4\times 10^{-17} \left( \frac{M_{pl}}{\Gamma}\right)^{1/4} M_{pl},
\end{eqnarray}
where $g_{reh}=106.75$ is the effective number of degrees of freedom for the Standard Model. So, taking into account the bound (\ref{bound}),
the reheating temperature ranges between  $7\times 10^3$ GeV and  $2\times 10^7$ GeV.

\

\subsection{Decay after the end of kination}

In the case  that the decay of the $\chi$-field is after the end of kination,
one  has to impose ${\Gamma}\leq H(\tau_{end})\equiv H_{end}$, where we have denoted by $\tau_{end}$ the time at which kination ends. Taking this into account, one has 
\begin{eqnarray}\label{31}
H^2_{end}=\frac{2\rho_{\varphi, end}}{3M_{pl}^2}\quad \mbox{and} \quad \rho_{\varphi, end}=\bar{\rho}_{\varphi}\left( \frac{\bar{a}}{a_{end}} \right)^6=
\frac{\langle \bar{\rho}\rangle^2}{\bar{\rho}_{\varphi}},
\end{eqnarray}
where we have used that the kination ends when ${\langle {\rho}(\tau_{end})\rangle}={{\rho}_{\varphi, end}}$, meaning 
$\left(\bar{a}/a_{end} \right)^3=
\frac{\langle\bar{\rho}\rangle}{\bar{\rho}_{\varphi}}$. So, the condition ${\Gamma}\leq H_{end}$ leads to the bound 
\begin{eqnarray}\label{bound1}
\Gamma\leq 10^{-23} M_{pl}.
\end{eqnarray}

\

 On the other hand, assuming once again instantaneous thermalization, the reheating temperature (i.e., the temperature of the universe when the thermalized plasma starts to dominate) will be obtained when all the superheavy particles decay, i.e. when $H\sim \Gamma$, obtaining
\begin{eqnarray}
T_{reh}=\left( \frac{30}{\pi^2 g_{reh}} \right)^{1/4}\langle\rho_{dec}\rangle^{1/4}= \left( \frac{90}{\pi^2 g_{reh}} \right)^{1/4}\sqrt{{\Gamma}M_{pl}}~,
\end{eqnarray}
where we have used that, after  the end of the kination regime, the energy density of the produced particles dominates the one  of the inflaton field. 
%Then, we will have 
%\begin{eqnarray}\label{decayafter}
%T_{reh}\cong 7\times 10^{-3} hg_*^{-1/4} M_{pl}.
%\end{eqnarray}

Consequently,  since the BBN epoch occurs at the $1$ MeV regime 
and taking once again  $g_{reh}=106.75$, one can find that, in that case,  the reheating temperature is bounded by
\begin{eqnarray}
1 \mbox{ MeV}\leq T_{reh} \leq 5\times 10^6 \mbox{ GeV }.
\end{eqnarray}

\

\begin{remark}
Since for $\alpha$-attractors in QI the number of produced particles is higher than in LQI, then the reheating temperature also increases. In fact, following step by step the same calculations that we have done above, we have obtained a very high reheating temperature when the decay is before the end of kination, between $10^7$ GeV and $10^9$ GeV. And, when the decay is after the end of kination, the reheating temperature belongs to the range between $1$ MeV and $10^9$ GeV.
\end{remark}

%{\color{red}
%Podriem calcular la energia per diferents masses $10^{13}$ $10^{11}$ $10^{9}$ i $10^7$ %GeV,
% veure les cotes per la temperatura de reescalfament, i fer una taula. 
%}

\section{Gravitational Waves and the BBN success}
\subsection{The overproduction of GWs}
\label{sec-overproduction}

In the seminal paper  \cite{pv},  the authors pointed out that in Quintessential Inflation a reheating due to the gravitational production of light particles is incompatible with the overproduction of Gravitational Waves. However, as we will show in this subsection,  for our LQI model, (and also in $\alpha$- QI, because in this scenario, as we have already explained, the production of particles is greater than in LQI),  when the reheating is due to the gravitational creation of superheavy particles, then the overproduction of GW's does not disturb  the BBN success.

\

To prove our statement, first of all we recall that the energy density of 
 the produced  GWs during the phase transition from the end of inflation to the beginning of kination is given by \cite{Giovannini99}
 \begin{eqnarray}
\langle\rho_{GW}(\tau)\rangle\cong \frac{H_{kin}^4}{2\pi^3}\left( \frac{a_{kin}}{a(\tau)}\right)^4\cong 
10^{-2} H_{kin}^4\left( \frac{a_{kin}}{a(\tau)}\right)^4.
\end{eqnarray}

Then, taking into account that
the success of the BBN demands that the ratio of the energy density of GWs to the one of the produced particles at the reheating time satisfies
\cite{dimopoulos}
\begin{eqnarray}\label{bbnconstraint}
\frac{\langle\rho_{GW, reh}\rangle}{\langle\rho_{reh}\rangle}\leq 10^{-2},
\end{eqnarray} 
we will see that the constraint (\ref{bbnconstraint}) is  overcome when the decay of the superheavy  particles is previous to the end of  kination. Effectively,
if the decay occurs before the end of kination one has 
\begin{eqnarray}
\frac{\langle\rho_{GW, reh}\rangle}{\langle\rho_{reh}\rangle}=
\frac{\langle\rho_{GW, dec}\rangle}{\langle\rho_{dec}\rangle},
\end{eqnarray} 
because the energy density of light relativistic particles evolves as the one of the GW's.
Next, noting  that during kination the energy density of the background scales as $a^{-6}$, it yields that 
\begin{eqnarray}
\left(\frac{a_{kin}}{a_{dec}} \right)^4=
\left(\frac{\rho_{\varphi,dec}}{\rho_{\varphi,kin}} \right)^{2/3},
\end{eqnarray} 
and thus, using our previous  results  (see formula (\ref{LQIrho})) and recalling  that the value of the Hubble rate at the beginning of kination is $H_{kin}\cong 4\times 10^{-8} M_{pl}$, we get 
\begin{eqnarray}
\frac{\langle\rho_{GW, reh}\rangle}{\langle\rho_{reh}\rangle}\cong 
 10^{-1} \left(\frac{\Gamma}{M_{pl}} \right)^{1/3}.
\end{eqnarray} 

Therefore, the bound (\ref{bbnconstraint}) is overcome when
\begin{eqnarray}
\Gamma\leq 7\times 10^{-4} M_{pl},
\end{eqnarray}
which is completely compatible with the bound (\ref{bound}).

\

On the other hand, when  the decay is produced after the end of kination, and assuming once again instantaneous thermalization,  the reheating time will coincide with the decay one. Then, since
$\langle\rho_{ dec}\rangle=3{\Gamma}^2M_{pl}^2$ and 
\begin{eqnarray}
H_{dec}=H_{end}\left( \frac{a_{end}}{a_{dec}} \right)^{3/2}\Longrightarrow \left( \frac{a_{end}}{a_{dec}} \right)^{3/2}=\sqrt{\frac{3}{2}}\frac{\Gamma M_{pl}\sqrt{\bar{\rho}_{\varphi}}}
{\langle \bar{\rho}\rangle},
\end{eqnarray}
where we have used that $H_{end}= \sqrt{\frac{2}{3}}
\frac{\langle \bar{\rho}\rangle}{M_{pl}\sqrt{\bar{\rho}_{\varphi}}}$.

Thus, we  will have 
\begin{align}
\langle\rho_{GW, dec}\rangle=\langle\rho_{GW, end}\rangle\left( \frac{a_{end}}{a_{dec}} \right)^4= 
\langle\rho_{GW, end}\rangle\left(  \sqrt{\frac{3}{2}}\frac{\Gamma M_{pl}\sqrt{\bar{\rho}_{\varphi}}}
{\langle \bar{\rho}\rangle}   \right)^{8/3}
%=10^{-2} H^4_{kin}\Theta^{-4/3}\left( \frac{\Gamma}{\sqrt{2}H_{kin} }  \right)^{8/3},
\end{align}
and, using that 
\begin{eqnarray}
\left(\frac{a_{kin}}{a_{end}}  \right)^4=  \frac{\rho_{\varphi, end}}{\rho_{\varphi, kin}} \quad \mbox{and} \quad 
\rho_{\varphi, end}=\frac{\langle \bar{\rho}\rangle^2}
{\bar{\rho}_{\varphi}},
\end{eqnarray}
we get
\begin{eqnarray}
\langle\rho_{GW, dec}\rangle=\left(\frac{2}{16}\right)^{1/3}\left( \frac{H_{kin}}{M_{pl}} \right)^2
\left(\frac{\bar{\rho}_{\varphi}}{\langle \bar{\rho}\rangle^2}\right)^{1/3}(\Gamma M_{pl})^{8/3}\cong 10^{-2}\Gamma^{8/3} M_{pl}^{4/3},
%{\bar{\rho}_{\varphi}}  \right)
\end{eqnarray}
and thus,
\begin{eqnarray}
\frac{\langle\rho_{GW, reh}\rangle}{\langle\rho_{reh}\rangle}=  \frac{\langle\rho_{GW, dec}\rangle}{\langle\rho_{dec}\rangle} 
\cong 3\times 10^{-3} \left(\frac{\Gamma}{M_{pl}} \right)^{2/3}\leq  10^{-17},
\end{eqnarray}
where we have used the bound (\ref{bound1}). So,  the constraint (\ref{bbnconstraint}) is clearly overcome.

\subsection{BBN constraints from the logarithmic spectrum of GWs}
\label{subsec-gw1}

As we have already explained,   GWs  are produced  in the post-inflationary period,  and its  logarithmic spectrum of GWs, namely
$\Omega_{GW}$ defined as $\Omega_{GW}\equiv \frac{1}{\rho_c}\frac{d\rho_{GW}(k)}{d\ln k }$ (where $\rho_{GW}(k)$ is the energy density spectrum of the produced GWs; $\rho_c=3H_0^2M_{pl}^2$, where $H_0$ is the present value of the Hubble rate, is the so-called {\it critical density}) scales as $k^2$ during kination \cite{rubio}, producing a spike in the spectrum of GWs at high frequencies. Then, so that GWs do not destabilize the BBN, the following bound must be imposed  (see Section 7.1 of \cite{maggiore}),
\begin{eqnarray}\label{integral}
I\equiv h_0^2\int_{k_{BBN}}^{k_{end}} \Omega_{GW}(k) d \ln k \leq 10^{-5},
\end{eqnarray}
where $h_0\cong 0.678$ parametrizes the experimental uncertainty to determine the current value of the Hubble constant and $k_{BBN}$, $k_{end}$ are the momenta associated to the horizon scale at the BBN and at the end of inflation respectively. As has been shown in \cite{Giovannini1}, the main contribution of the integral \eqref{integral} comes from the modes that leave the Hubble radius before the end of the inflationary epoch and finally re-enter during  { kination}, that means, for $k_{end}\leq k\leq k_{kin}$, where
$k_{end}=a_{end}H_{end}$ and $k_{kin}=a_{kin}H_{kin}$.  For these modes one can calculate the  logarithmic spectrum of GWs as in \cite{Giovannini} (see also \cite{rubio, Giovannini2, Giovannini3,Giovannini:2016vkr}),

\begin{eqnarray}\label{Omega}
\Omega_{GW}(k)=\tilde{\epsilon}\Omega_{\gamma}h^2_{GW} \left(\frac{k}{k_{end}}  \right)\ln^2\left(\frac{k}{k_{kin}}  \right),
\end{eqnarray}
where $h^2_{GW}=\frac{1}{8\pi}\left(\frac{H_{kin}}{M_{pl}}  \right)^2$
is the amplitude of the GWs; $\Omega_{\gamma}\cong 2.6\times 10^{-5} h_0^{-2}$ is the present density fraction of radiation, and the quantity $\tilde{\epsilon}$, which is approximately equal to $0.05$ for the Standard Model of particle physics,  takes into account the variation of massless degrees of freedom between decoupling and thermalization (see \cite{rubio, Giovannini1} for more details). As has been derived in \cite{Giovannini1}, the specific form of the expression above comes from the behavior of the Hankel functions for small arguments. Now, plugging expression (\ref{Omega}) into (\ref{integral}) and disregarding the sub-leading logarithmic terms, one finds  
\begin{eqnarray}\label{constraintx}
 2\tilde{\epsilon}h_0^2\Omega_{\gamma}h^2_{GW}\left( \frac{k_{kin}}{k_{end}} \right)\leq 10^{-5}
 \Longrightarrow 
 10^{-2}\left( \frac{H_{kin}}{M_{pl}} \right)^2\left( \frac{k_{kin}}{k_{end}} \right)\leq 1
 \Longrightarrow 
 10^{-17}\left( \frac{k_{kin}}{k_{end}} \right)\leq 1,  \end{eqnarray}
  because in our LQI model at the beginning of kination the Hubble rate is   $H_{kin}\sim 4\times 10^{-8} M_{pl}$, and the same happens for our $\alpha$-QI model.

  \

 So, we continue with the LQI, but  taking into account that all the reasoning is also valid for the other model. Then,  
to calculate the ratio $k_{kin}/k_{end} $,  we  will have to study the following
 two different situations:

\begin{enumerate}

\item When the decay occurs before the end of kination.

In this case the reheating  time coincides with the end of kination, and thus,
a simple calculation leads to 
\begin{eqnarray}
\frac{k_{kin}}{k_{end}} =\frac{k_{kin}}{k_{reh}}.
\end{eqnarray}

Using  the formulas
\begin{eqnarray}
\rho_{\varphi, reh}=\rho_{\varphi, kin}\left(\frac{a_{kin}}{a_{reh}}\right)^6\qquad \mbox{and}\qquad
\langle\rho_{ reh}\rangle=\langle \rho_{ dec}\rangle\left(\frac{a_{dec}}{a_{reh}}\right)^4,
\end{eqnarray}
we get 
\begin{eqnarray}
\frac{a_{kin}}{a_{reh}}=\left( \frac{\langle\rho_{ dec}\rangle}{\rho_{\varphi, kin}} \right)^{1/6}
\left(\frac{a_{dec}}{a_{reh}}\right)^{2/3}=\left( \frac{\langle\rho_{dec}\rangle}{\rho_{\varphi, kin}} \right)^{1/6}\left( \frac{\langle\rho_{ dec}\rangle}{\rho_{\varphi, dec}} \right)^{1/3},
\end{eqnarray}
where we have used the relation $\left(\frac{a_{dec}}{a_{reh}}\right)^2=
\frac{\langle\rho_{dec}\rangle}{\rho_{\varphi, dec}}$. Then, taking into account that $H_{reh}=\Gamma \left(\frac{a_{dec}}{a_{reh}}\right)^3$, we obtain
\begin{eqnarray}
\frac{k_{kin}}{k_{end}} =\frac{k_{kin}}{k_{reh}}=
\frac{H_{kin}a_{kin}}{H_{reh}a_{reh}}=\frac{H_{kin}}{\Gamma}\frac{\rho_{\varphi, dec}}{\langle\rho_{dec}\rangle}
\left( \frac{\rho_{\varphi, dec}}{\rho_{\varphi, kin}} \right)^{1/6}.
\end{eqnarray}

\

 Finally, from our previous results (\ref{LQIrho})
  \begin{eqnarray}
  \rho_{\varphi, dec}=3\Gamma^2M_{pl}^2, \quad \langle\rho_{dec}\rangle\sim  10^{-21}\Gamma M_{pl}^3 \quad \mbox{and} \quad  \rho_{\varphi, kin}=3H_{kin}^2M_{pl}^2,
  \end{eqnarray}
we arrive at
\begin{eqnarray}
\frac{k_{kin}}{k_{end}}\sim 10^{15}\left(\frac{\Gamma}{M_{pl}} \right)^{1/3},
\end{eqnarray}
and consequently the constraint (\ref{constraintx}) becomes 
\begin{eqnarray}
\left(\frac{\Gamma}{M_{pl}} \right)^{1/3}\leq 10^2,
\end{eqnarray}
which is obviously overcome for all the viable values of $\Gamma$, i.e, for all values between $10^{-21} M_{pl}$ and $10^{-8} M_{pl}$.

\item When the  decay occurs  after the end of kination.

Now, since during kination we have $H_{end}=H_{kin}\left(\frac{a_{kin}}{a_{end}}\right)^3$, one gets
\begin{eqnarray}
\frac{k_{kin}}{k_{end}}=\left(\frac{H_{kin}}{H_{end}} \right)^{2/3},
\end{eqnarray}
and taking into account that
\begin{eqnarray}
H_{end}=\sqrt{\frac{2\langle\bar{\rho}\rangle^2}{3M_{pl}^2\bar{\rho}_{\varphi}}}\sim 3\times 10^{-22}M_{pl}\qquad \mbox{and} \qquad H_{kin}\sim 4\times 10^{-8} M_{pl}, 
\end{eqnarray}
the bound (\ref{constraintx}) is completely overcome.

\end{enumerate}

\section{Conclusions}

In the present work we have numerically studied the gravitational particle production of superheavy particles conformally coupled to gravity for two classes of QI scenarios, namely LQI and $\alpha$-QI. To calculate  the energy density of the produced particles we have used the well-known diagonalisation method, where the key point is the calculation and interpretation  of the time-dependent 
$\beta$-Bogoliubov coefficient. 

\

In fact, this coefficient encodes all the polarization effects (creation and annihilation of pairs named, in the Russian literature,  {\it quasiparticles}) and the produced superheavy particles -the real ones which after decaying into lighter ones form a relativistic plasma which reheats the universe-  during the phase transition from the end of inflation to the beginning of kination. However, and this is the main observation of the present work, the polarization effects disappear soon after the end of the phase transition (during the kination regime), i.e., when the evolution of the universe gets adiabatic again. So, we have numerically checked  that effectively the value of the $\beta$-Bogoliubov coefficient stabilizes during the kination, which allows us to compute it numerically, and thus, to calculate numerically the energy density of the  particles created during this phase transition.

\

Once these superheavy particles have been created, they must decay into lighter ones to form a relativistic plasma which eventually becomes dominant and matches with the hot Big Bang universe. Then two different situations arise, namely when the decay occurs  before the end of kination regime and when the decay occurs after the end of the kination regime. Thus,  since we have  numerically computed the energy density of the superheavy particles, 
for both situations we have been able to  calculate the reheating temperature of the universe, which depends on the decay rate of these superheavy particles and whose maximum value is quite big -more or less around $10^7$ GeV in LQI and around $10^9$ GeV in $\alpha$-QI-, which demystifies the belief, never checked numerically,  that heavy masses suppress the particle production, thus leading to  an abnormally  low reheating temperature. In fact, the main contribution of particle production is to a long wavelength regime, which is impossible to quantify analytically because only ultraviolet effects can be calculated with analytic methods, and this is the reason why   in many papers the production of superheavy particles is simply disregarded.

\

Finally, we have checked that  the 
 overproduction of GWs does not distrub   the BBN, because all the bounds preserving  its success are clearly overcome.

\section*{Acknowledgments}

The investigation of JdH has been supported by MINECO (Spain) grant   MTM2017-84214-C2-1-P, and  in part by the Catalan Government 2017-SGR-247. L.A.S thanks the School of Mathematical Sciences (Queen Mary University of London) for the support provided.
%SP acknowledges the research grant under Faculty Research and Professional Development Fund (FRPDF) Scheme of Presidency University, Kolkata, India.  The authors thank Prof. M. Giovannini and Prof. J. D. Barrow for useful correspondence. 

\end{document}